\documentclass{aa}
\usepackage{graphics}
\usepackage{times}
\begin{document}
\thesaurus{12(12.03.4; 12.04.3; 11.04.1; 
11.07.1; 11.11.1; 11.12.1)}
\title{
Investigations of the Local Supercluster Velocity Field
}
\subtitle{
II. A study using Tolman-Bondi solution and galaxies with accurate 
distances from the Cepheid $PL$-relation}
\author{
Ekholm,~T.\inst{1,2}
\and
Lanoix,~P.\inst{1}
\and
Teerikorpi,~P.\inst{2}
\and
Paturel,~G.\inst{1}
\and
Fouqu\'e,~P.\inst{3}
}
\offprints{
T.~Ekholm
}
\institute{
CRAL - Observatoire de Lyon,
F69561 Saint Genis Laval CEDEX, France
\and
Tuorla Observatory,
FIN-21500 Piikki\"o,
Finland
\and
ESO, Santiago, Chile
}
\date{received, accepted}
\maketitle
%
%
%
%
\begin{abstract}
A sample of 32 galaxies with accurate distance moduli
from the Cepheid $PL$-relation (Lanoix \cite{Lanoix99}) 
has been used to study the dynamical
behaviour of the Local (Virgo) supercluster. We used analytical
Tolman-Bondi (TB) solutions for a spherically symmetric density excess
embedded in the Einstein-deSitter universe ($q_0=0.5$). Using 12
galaxies within $\Theta=30\degr$ from the centre we found a mass
estimate of $1.62M_\mathrm{virial}$ for the Virgo cluster. This
agrees with the finding of Teerikorpi et al. (\cite{Teerikorpi92})
that TB-estimate may be larger than virial mass estimate
from Tully \& Shaya (\cite{Tully84}). 
Our conclusions do not critically depend on our 
primary choice of the global
$H_0=57\mathrm{\ km\, s^{-1}\, Mpc^{-1}}$ established from
SNe Ia (Lanoix \cite{Lanoix99}). 
The remaining galaxies outside Virgo region do not disagree with
this value.
Finally, we also found a TB-solution
with the $H_0$ and $q_0$ cited yielding exactly one virial
mass for the Virgo cluster.
\keywords{Cosmology: theory -- distance scale --
Galaxies: distances and redshifts -- Galaxies: general --
Galaxies: kinematics and dynamics -- Local Group} 
\end{abstract}
%
%
%
%
\section{Introduction}
When studying the dynamical properties of a structure like
the Local (Virgo) supercluster (LSC) one obviously cannot use the standard
solutions to the Friedman-Robertson-Walker (FRW) world 
model which are valid in a homogeneous
environment only. General Relativity does, however, provide
means for this kind of a problem, namely the so-called Tolman-Bondi (TB)
model. While no longer requiring strict homogeneity,
spherical symmetry is a necessary requisite. 
In the present paper we address the question how well the
velocity field in the LSC follows the
TB-prediction, i.e. how well it meets the requirement
of spherical symmetry. 

Previous studies 
(Tully \& Shaya \cite{Tully84}; Teerikorpi et al.
\cite{Teerikorpi92} - hereafter Paper I) provided evidence
for the expected velocity field if the mass of the Virgo cluster
itself is roughly equal to its virial mass. However, the 
Tully-Fisher distances used caused significant scatter. Now we possess
a high-quality sample of galaxies with accurate distances from
Cepheids. 

Tolman (\cite{Tolman34}) found the general solution to the
Einstein's field equations for a spherically symmetric
pressure-free dust universe in terms of the comoving coordinates.
The metric can be expressed as:
%
\begin{equation}
\label{E1}
\\ ds(r,\tau)^2=d\tau^2-\frac{R'(r,\tau)^2}{1+f(r)}dr^2-R(r,\tau)^2d\Omega^2,
\end{equation}
where $d\Omega^2=d\theta^2+\sin^2\theta d\phi^2$ and $f(r)$
is some unknown function of the comoving radius $r$.
$R$ corresponds to the usual concept of distance.
Integration of the equation of motion yields:
%
\begin{equation}
\label{E2}
\\ \dot{R}^2=\frac{F(r)}{R}+f(r).
\end{equation}
$\dot{R}$ refers to $\partial R/\partial \tau$ and $R'$ to
$\partial R/\partial r$. $F(r)$ is another arbitrary
function of $r$.

Bondi (\cite{Bondi47}) interpreted Eq.~\ref{E2} as the total energy
equation: $f(r)$ is proportional to the total energy of the
hypersurface of radius $r$ and $F(r)$ is proportional to the
mass inside $r$. In this interpretation it is clear that one must
require $F(r)>0$. Eq.~\ref{E2} integrates into three distinctive
solutions. Two of them are expressed in terms of a 
parameter $\eta$, the development angle of
the mass shell in consideration:
%
\begin{eqnarray}
\label{E3}
R &=& \frac{F}{2f} (\cosh\eta-1) \nonumber \\
\tau-\tau_0(r) &=& \frac{F}{2f^{3/2}}(\sinh\eta-\eta),
\end{eqnarray}
for positive energy function $f(r)>0$,
%
\begin{eqnarray}
\label{E4}
R &=& \frac{F}{-2f} (1-\cos\eta) \nonumber \\
\tau-\tau_0(r) &=& \frac{F}{2(-f^{3/2})}(\eta-\sin\eta),
\end{eqnarray}
for negative energy function $f(r)<0$.
The third solution is for $f(r)=0$: 
%
\begin{equation}
\label{E5}
\\ R = \left(\frac{9F}{4}\right)^{1/3}\left[\tau-\tau_0(r)\right]^{2/3}.
\end{equation}
It is important to note that the solution has {\it three} undefined
functions: $f(r)$, $F(r)$ and $\tau_0(r)$. One of these can be
fixed by defining some arbitrary transformation of the comoving
coordinate $r$. In our application we set $\tau_0(r)\equiv0$ and
interpret $\tau$ as time elapsed since the Big Bang, i.e. we equal
$\tau$ with the age of the Universe $T_0$.

To find a TB-prediction for the
observed velocity, we use the formulation of 
Paper I, based on the solutions given by
Olson \& Silk (\cite{Olson79}). 
In this approach the development angle
$\eta$ needed for the velocity is solved with the help of
a function:
%
\begin{equation}
\label{E6}
\\ A(R,T_0) = \frac{\sqrt{GM(R)}T_0}{R^{3/2}},
\end{equation}
where $R$ is the distance of spherical mass shell from the origin of the
TB-metric, $M(R)$ is the mass contained within the shell and $G$
is the gravitational constant. $\eta$ is solved either from Eq.~7 or
Eq.~8 of Paper I depending on value
$A(R,T_0)$. Ekholm (\cite{Ekholm96}; hereafter E96) 
calculated in his Appendix B the
exact value of $A$ dividing the family of solutions into open, hyperbolic
solutions ($f(r)>0$) and closed, elliptic solutions ($f(r)<0$):
$A=\sqrt2/3\approx0.47$.
%
%
%
%
\section{Details of the model used}
In Paper I, Ekholm \& Teerikorpi
(\cite{Ekholm94}; hereafter ET94) and E96 one chose as a starting
point a ``known" mass $M(R)$ for each $R$ inferred from some
density distribution given as a sum of an excess density and
the uniform cosmological background. 
We assume:
%
\begin{equation}
\label{E7}
\\ \rho(R) = \frac {3H_0^2q_0}{4\pi G}(1+kR^{-\alpha}).
\end{equation}
$H_0$ is the Hubble constant, $q_0$ is the deceleration parameter.
$k$ and $\alpha$ define the details of the density model and will
be explained shortly. 

E96 further developed this formalism by
defining a ``local Hubble constant" $H_0^*=V_\mathrm{cosm}(d\!=\!\!1)$,
where $d$ is calculated from Eq.~\ref{E14}.
By setting $d=1$ we can fix the mass excess $k'$ independently
of the density gradient $\alpha$.
In this manner
the quantity $A$ can be expressed in terms of $d$ and $q_0$
(Eq.~7 in E96):
%
\begin{equation}
\label{E8}
\\ A(d,q_0)=C(q_0)\sqrt{q_0(1+k'd^{-\alpha})}.
\end{equation}
The factor $C(q_0)$ depends on the cosmological FRW world model chosen
(cf. Eqs.~8-10 in E96). 
The mass contained by a shell of radius $d$ is
(Eq.~4 in E96):
%
\begin{equation}
\label{E9}
\\ M(d)=\frac{H_0^2}{G}q_0R_\mathrm{Virgo}^3d^3(1+k'd^{-\alpha}).
\end{equation}
$k'$ ($k$ normalized to the Virgo distance) provides the mass
excess within a sphere of radius $d=1$ and can be fixed from
our knowledge of the observed velocity of the centre of the mass
structure and of the infall velocity of the Local Group (LG) as
follows. Let $V_\mathrm{Virgo}^\mathrm{obs}$ be the observed
velocity of Virgo and $V_\mathrm{LG}^\mathrm{in}$ the infall 
velocity of the LG. The predicted velocity of a galaxy with
respect to the centre of the structure in consideration was given by
E96 (his Eq.~11):
%
\begin{equation}
\label{E10}
\\ v(d) = \frac{\gamma d V_\mathrm{Virgo}^\mathrm{obs}\phi(\eta_0)}{C(q_0)}
\end{equation}
with
%
\begin{equation}
\label{E11}
\\ \gamma = 
(V_\mathrm{Virgo}^\mathrm{obs}+V_\mathrm{LG}^\mathrm{in})
/V_\mathrm{Virgo}^\mathrm{obs}.
\end{equation}
$\phi(\eta_0)$ is the angular part of Eqs.~15 and~16 given by
ET94, where $\eta_0$ is the
development angle corresponding to the given value of $A$. 
Hence for each $V_\mathrm{LG}^\mathrm{in}$ one may fix $k'$ by
requiring:
%
\begin{equation}
\label{E12}
\\ v(1) = V_\mathrm{Virgo}^\mathrm{obs}.
\end{equation}
One clear advantage
of this formulation is -- being perhaps otherwise idealistic -- that
the value of $k'$ depends only on the two velocities given above
for a fixed FRW world model ($H_0$, $q_0$).
Thus $\alpha$ will alter only the distribution of dynamical matter
(the larger the $\alpha$ the more concentrated the structure).

Finally, we need the predicted counterpart of the observed velocity
of a galaxy. In the present application it is solved from:
%
\begin{equation}
\label{E13}
\\ V_\mathrm{pred}(d_\mathrm{gal}) = 
V_\mathrm{Virgo}^\mathrm{obs}\cos\Theta\pm
v(d)\sqrt{1-\sin^2\Theta/d^2},
\end{equation}
where $d_\mathrm{gal}=R_\mathrm{gal}/R_\mathrm{Virgo}$ 
is the relative distance of a galaxy from the LG,
$\Theta$ is the corresponding angular distance and $d$, the
distance to the galaxy measured from the centre of the structure,
is evaluated from:
%
\begin{equation}
\label{E14}
\\ d = \sqrt{1+d_\mathrm{gal}^2-2d_\mathrm{gal}\cos\Theta}.
\end{equation}
The ($-$) -sign is valid for points closer than the tangential point
$d_\mathrm{gal}<\cos\Theta$ and ($+$) -sign for 
$d_\mathrm{gal}\ge\cos\Theta$.
%
%
%
%
\section{The sample}
Eq.~\ref{E14} reveals one significant difficulty in an analysis
of this kind. In order to find the relative distances $d_\mathrm{gal}$ 
one needs good estimates for the distances to the
galaxies $R_\mathrm{gal}$. Such are difficult to obtain. Distances
inferred from photometric data via, say, the Tully-Fisher relation
are hampered by large scatter and as a result, by the Malmquist 
bias which is difficult to correct for. 
Distances inferred from
velocities using the Hubble law are obviously quite unsuitable. 

The HIPPARCOS satellite has provided a sample of galactic Cepheids
from which 
Lanoix et al. (\cite{Lanoix99a})
obtained a new calibration of the $PL$-relation
in both V and I band. 
Lanoix (\cite{Lanoix99})
extracted a sample of 32 galaxies from the
Extragalactic Cepheid Database (Lanoix et al. \cite{Lanoix99b}) 
and, by taking into account the incompleteness bias in the extragalactic
$PL$-relation (Lanoix et al. \cite{Lanoix99c}), he inferred the
distance moduli for these galaxies  
with 23 based on HST measurements and
the rest on groundbased measurements. It is also important to
note that the sample is very homogeneous and the method for calculating
the distance modulus is quite accurate.  
The photometric distances found in
this way are far more accurate and of better quality 
than those derived using Tully-Fisher
relation, a fact compensating the smallness of our sample.

We intend to use these galaxies together
with the TB-model described in the previous two sections
to study the gross features of the dynamical structure of the
LSC. Because we need only the observed velocities,
$V_\mathrm{obs}$ and the Cepheid distance moduli $\mu$, we can
avoid the usual difficulties and caveats of other photometric distance
determinations. The velocities were extracted from the Lyon-Meudon
Extragalactic Database LEDA. By observed velocity we refer to the
mean heliocentric velocity corrected to the LG centroid
according to Yahil et al. (\cite{Yahil77}). 

We need, however, some additional information: 
$V_\mathrm{Virgo}^\mathrm{obs}$, $V_\mathrm{LG}^\mathrm{in}$ and
$R_\mathrm{Virgo}$. For the observed velocity of the centre of the
LSC the value preferred by Paper I is
$V_\mathrm{Virgo}^\mathrm{obs}=980.0\mathrm{\ km\,s^{-1}}$. For the
infall velocity of the LG into the centre we choose
$V_\mathrm{LG}^\mathrm{in}=220\mathrm{\ km\,s^{-1}}$
(Tammann \& Sandage \cite{Tammann85}). It is worth noting that though
this value has fluctuated, this relatively old value is still
quite compatible with recent re-evaluations (cf. Federspiel et al.
\cite{Federspiel98}). Furthermore, the fluctuations have not been
very significant. We also need an estimate for the distance of
the centre of the LSC, $R_\mathrm{Virgo}$. One possibility
is to establish the Hubble constant $H_0$ from some independent
method. Lanoix (\cite{Lanoix99}) found using SNe Ia's:\footnote{ 
$H_0$ derived is not completely independent because both the SNe
and the extragalactic $PL$-relation are calibrated with same
Cepheids.}
%
\begin{equation}
\label{E15}
\\ H_0 = 57\pm3\mathrm{\ km\,s^{-1}\,Mpc^{-1}}.
\end{equation}
This {\it global} value of $H_0$ is in good agreement with our
more local results using both the direct and inverse Tully-Fisher relations
(Theureau et al. \cite{Theureau97}; Ekholm et al. \cite{Ekholm99}). 
$H_0$ and the given velocities yield 
$R_\mathrm{Virgo}=21.0\mathrm{\ Mpc}$, which is in good agreement
with $R_\mathrm{Virgo}=20.7\mathrm{\ Mpc}$ found by 
Federspiel et al. (\cite{Federspiel98}). Note also that Federspiel
et al. (\cite{Federspiel98}) found the same value for $H_0$ from relative
cluster distances to Virgo. Finally, one should recognize that
our distance estimate is valid only if Virgo is at rest with
respect to the FRW-frame. If not, the cosmological velocity of
Virgo is something more complicated than simply the sum of
$V_\mathrm{LG}^\mathrm{in}$ and $V_\mathrm{Virgo}^\mathrm{obs}$.
%
%
%
%
\section{Results in the direction of Virgo ($\Theta<30\degr$)}
\begin{figure}
\resizebox{\hsize}{!}{\includegraphics{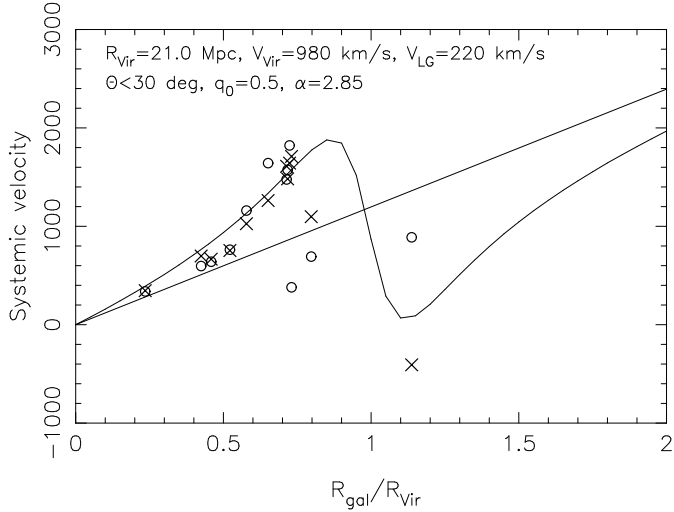}}
\caption{The systemic velocity vs. distance for galaxies listed 
in Table 1 for the Model 1.
Circles refer to the observed velocities and crosses to the
TB-predictions. The straight line is the Hubble law for
$H_0=57\mathrm{\ km\, s^{-1}\, Mpc^{-1}}$ and the
curve is the theoretical TB-pattern for $\Theta=7\degr$.
Note that the most discrepant galaxies (\object{NGC 4639}
and \object{NGC 4548}) are actually closest to the Virgo
centre in the sky.
}
\label{F1}
\end{figure}

We restrict ourselves to the Einstein-deSitter universe by
assigning $q_0=0.5$.  
As regards Eq.~\ref{E8} we now possess the distance $d$ for
each galaxy as well as the value $k'=0.606$ from Eq.~\ref{E12}. 
We also have $\gamma$ from Eq.~\ref{E11}. 
We need to estimate the best value for
$\alpha$. 
We devised a simple statistical test by finding which value
or values of $\alpha$ minimize the average 
$\vert V_\mathrm{obs}-V_\mathrm{pred}\vert$
of sample galaxies in the Virgo direction. We found minimum values
around $\alpha=2.7-3.0$.
We adopt $\alpha=2.85$.
The resulting systemic velocity\footnote{Systemic velocity is a
combination of the cosmological velocity and the
velocity induced by Virgo. This is in our case the observed velocity
defined in Sect.~4. It could contain also other components, but
we assume that the Virgo infall dominates.} 
vs. distance diagram is shown in
Fig~\ref{F1}. 
The observed velocities are labelled with circles and the predicted
values with crosses. The straight line is what one expects from
Hubble law with our choice of $H_0$ and the curve is the theoretical
TB-solution for $\Theta=7\degr$ (most of the galaxies
have small angular distances)\footnote{
This curve is simply to guide the eye. The actual comparison
is made between each observed and predicted point.}.
The relative distances and predicted velocities are also
tabulated in Table 1 at columns 4 and 5 for this Model 1. 
The galaxies follow the overall TB-pattern, with velocities
steeply increasing when the Virgo center is approached.
Exceptions are \object{NGC 4639} lying at 1.14 
and \object{NGC 4548} at 0.73.
%
\begin{table}
\begin{center}
\begin{tabular}{lccccccc}
\hline
 Name & $\Theta$ & $V_\mathrm{obs}$ & $d_{21}$  & $V_\mathrm{pred,\ 1}$ &
$d_{20.7}$ & $V_\mathrm{pred,\ 2}$ \\
\hline
 \object{IC 4182}  & 26.4 &  337.0 & 0.24 &  345.0 & 0.24 &  335.0 \\
 \object{NGC 3351}  & 26.2 &  640.0 & 0.46 &  666.0 & 0.47 &  545.0 \\
 \object{NGC 3368}  & 25.5 &  761.0 & 0.52 &  753.0 & 0.53 &  728.0 \\
 \object{NGC 3627}  & 17.3 &  596.0 & 0.42 &  698.0 & 0.43 &  680.0 \\
 \object{NGC 4321}  &  4.0 & 1475.0 & 0.71 & 1611.0 & 0.72 & 1593.0 \\
 \object{NGC 4414}  & 18.9 &  692.0 & 0.80 & 1098.0 & 0.81 & 1040.0 \\
 \object{NGC 4496A} &  8.4 & 1574.0 & 0.72 & 1482.0 & 0.73 & 1453.0 \\
 \object{NGC 4535}  &  4.3 & 1821.0 & 0.72 & 1642.0 & 0.73 & 1623.0 \\
 \object{NGC 4536}  & 10.2 & 1641.0 & 0.65 & 1261.0 & 0.66 & 1235.0 \\
 \object{NGC 4548}  &  2.4 &  379.0 & 0.73 & 1710.0 & 0.74 & 1695.0 \\
 \object{NGC 4639}  &  3.0 &  888.0 & 1.14 & -409.0 & 1.15 & -351.0 \\
 \object{NGC 4725}  & 13.9 & 1160.0 & 0.58 & 1025.0 & 0.59 & 1001.0 \\
\hline
\end{tabular}
\end{center}
\caption{ 
Parameters for the 12 galaxies within $\Theta<30\degr$ from the
adopted centre of the LSC.
Column 1: Name,
Column 2: Angular distance $\Theta$,
Column 3: Observed velocity $V_\mathrm{obs}$,
Column 4: Relative distance with $R_\mathrm{Virgo}=21\mathrm{\ Mpc}$,
Column 5: Predicted velocity from Model 1, 
Column 6: Relative distance with $R_\mathrm{Virgo}=20.7\mathrm{\ Mpc}$, and 
Column 7: Predicted velocity from Model 2. 
}
\end{table}

\object{NGC 4639} has a positive
velocity when -- according to the model -- it should be falling into
Virgo from the backside. Can we explain this strange behaviour?
We note it is at a small angular distance from
Virgo ($\Theta=3.0\degr$). 
Perhaps \object{NGC 4639} has fallen through
the centre and still has some of its frontside infall velocity
left. There are, however, no 
reports of significant hydrogen deficiency which one would expect
if the galaxy actually had travelled through the centre.
Maybe this galaxy is a genuine member of the Virgo cluster
thus having a rather large dispersion in velocity.
In fact, Federspiel et al. 
(\cite{Federspiel98}) include it to their ``fiducial sample" of
Virgo cluster galaxies belonging to the subgroup ``big A" around
\object{M87}. As a matter of fact our chosen centre ($l=284\degr$,
$b=74.5\degr$) lies close to the massive pair \object{M86/87}. 

\object{NGC 4548} is also close to Virgo and has a small angular distance
($\Theta=2.4\degr$). Now, however the situation is reversed. This
galaxy should be falling into Virgo with a high positive velocity
but has a small velocity. Paper I classifies this galaxy belonging
to a region interpreted as a component expanding from Virgo
(cf. Fig.~8 (Region B) and Table~1c in Paper I).

Finally, because both galaxies are close to Virgo at small angular
distances even a slight error in distance determination could
result in a considerable error in velocities. Hence both galaxies
might follow the TB-curve but error in distance has
distorted the figure.
 
%
We also tested the exact values given by 
Federspiel et al. (\cite{Federspiel98}) with $\alpha=2.85$.
Now $V_\mathrm{Virgo}^\mathrm{obs}=920\mathrm{\ km\,s^{-1}}$ and
$R_\mathrm{Virgo}=20.7\mathrm{\ Mpc}$ yield with the same infall
velocity $H_0=55\mathrm{\ km\,s^{-1}\,Mpc^{-1}}$. We found
$k'=0.641$. 
The behaviour of the systemic velocity as a function of distance is quite
similar to Fig.~\ref{F1}.  
This solution is named as Model 2 and given 
in columns 6 and 7 in Table 1.
Due to the paucity of the sample it is not possible to
decide between these models.

Though we have only a few points, it is quite remarkable how well
the dynamical behaviour of galaxies
in the direction of Virgo is demonstrated by a simple model. That the galaxies
follow so well a spherically symmetric model despite the observed
clumpiness of the Virgo region is promising. 
Fig.~\ref{F1} clearly
lends credence to a presumption that the gravitating matter 
is distributed more symmetrically than
the luminous matter. We have given an affirmative answer
to the question asked in the introduction. 
%
%
%
%
\section{Results outside Virgo ($\Theta>30\degr$)}
\subsection{On the chosen Hubble constant}
%
%
%
\begin{table}
\begin{center}
\begin{tabular}{lccccc}
\hline
 Name & $\Theta$ & $V_\mathrm{obs}$ & $R_\mathrm{Mpc}$ \\ 
\hline
 \object{IC 1613}  &  163.2 &  -62.0 &  0.69 \\
 \object{LMC}      &  107.4 &   82.0 &  0.05 \\
 \object{NGC 1365} &  132.7 & 1563.0 & 18.20 \\
 \object{NGC 2090} &  106.0 &  754.0 & 11.48 \\
 \object{NGC 224}  &  126.3 &  -13.0 &  0.87 \\
 \object{NGC 2541} &   63.8 &  645.0 & 11.59 \\
 \object{NGC 300}  &  154.1 &  125.0 &  2.17 \\
 \object{NGC 3031} &   61.8 &  124.0 &  3.37 \\
 \object{NGC 3109} &   52.7 &  130.0 &  1.02 \\
 \object{NGC 3621} &   48.4 &  436.0 &  6.61 \\
 \object{NGC 3198} &   43.3 &  704.0 & 13.68 \\
 \object{NGC 4603} &   53.4 & 2300.0 & 32.51 \\
 \object{NGC 5253} &   47.0 &  155.0 &  3.16 \\
 \object{NGC 5457} &   45.7 &  361.0 &  6.92 \\
 \object{NGC 598}  &  134.4 &   68.0 &  0.79 \\
 \object{NGC 6822} &  110.7 &    8.0 &  0.45 \\
 \object{NGC 7331} &  125.9 & 1115.0 & 14.39 \\
 \object{NGC 925}  &  126.4 &  782.0 &  8.87 \\
 \object{SEXA}     &   38.8 &  118.0 &  1.45 \\
 \object{SEXB}     &   37.9 &  139.0 &  1.39 \\
\hline
\end{tabular}
\end{center}
\caption{ 
Parameters for the 20 galaxies with $\Theta>30\degr$ from the
adopted centre of the LSC.
Column 1: Name,
Column 2: Angular distance $\Theta$,
Column 3: Observed velocity $V_\mathrm{obs}$, and
Column 4: Distance in Mpc. 
}
\end{table}

\begin{figure}
\resizebox{\hsize}{!}{\includegraphics{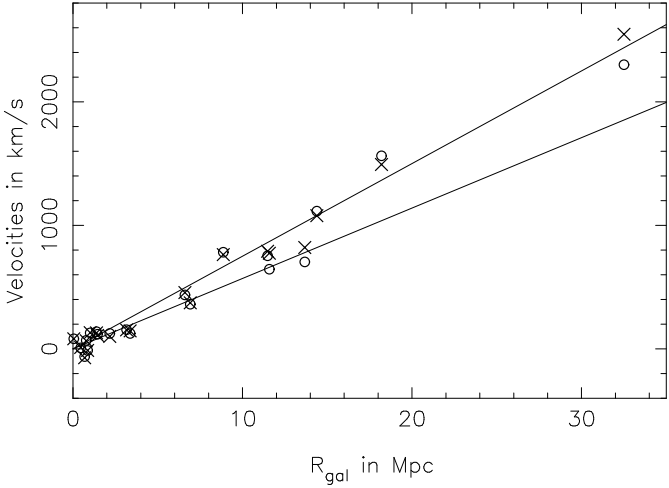}}
\caption{The observed systemic velocity (open circles)
vs. absolute distance for galaxies
outside the Virgo region. 
The straight lines are the predictions
for
$H_0=57\mathrm{\ km\, s^{-1}\, Mpc^{-1}}$ from SNe Ia 
and for
$H_0=75\mathrm{\ km\, s^{-1}\, Mpc^{-1}}$. Corrected velocities
are also plotted as crosses. Note that \object{NGC 4603} has a
very bad $PL$ relation in V and hence one should assign a low
weight to it. 
}
\label{F2}
\end{figure}

In Paper I the value of the Hubble constant was taken to be
$H_0=70\mathrm{\ km\,s^{-1}\,Mpc^{-1}}$, 
consistent with the Tully-Fisher calibration adopted at
the time and with the used Virgo distances (16.5 and 18.4 Mpc
giving different infall velocities of LG). In the present paper
we have {\it fixed} $H_0$ from more global considerations as
well as $V_\mathrm{LG}^\mathrm{in}$ corresponding to
$R_\mathrm{Virgo}=21\mathrm{\ Mpc}$.
As discussed by ET94 once the
velocity of the LG is fixed the predicted velocities no
longer depend on the value of $H_0$. It is still interesting to
see whether the sample galaxies locally agree with
$H_0=57\mathrm{\ km\, s^{-1}\, Mpc^{-1}}$.
This is done in Fig.~\ref{F2}. where we
have plotted the observed 
systemic velocities (open circles) outside
the Virgo region ($\Theta>30\degr$) as a function of the absolute
distance in Mpc.
The data are given in Table 2.
The galaxies follow quite
well the quiescent flow. In particular, the closest-by galaxies
($R_\mathrm{gal}<4\mathrm{\ Mpc}$) follow the linear 
prediction with surprising accuracy. 
Note also that galaxies listed in Table 1 (galaxies partaking in the
Virgo infall) would predict 
$H_0=76\pm9\mathrm{\ km\, s^{-1}\, Mpc^{-1}}$.
This too high a value clearly underlines the need for correct 
and adequate kinematical
model in Virgo region. 

However, two of the galaxies are clearly discrepant
(\object{NGC 1365} and \object{NGC 4603}) and two (\object{NGC 925} and
\object{NGC 7331}) disagree. 
First of all, \object{NGC 4603} is a distant galaxy
and the distance determination may be biased by the Cepheid
incompleteness effect (Lanoix et al. \cite{Lanoix99c}).
It also does not have a good $PL$-relation in V, which means
that one should assign a very low weight for it. 
On the
other hand, its velocity differs from the quiescent Hubble flow
only by $375\mathrm{\ km\, s^{-1}}$ which actually is not very
much considering the distance. It belongs to the Virgo Southern 
extension and its velocity could be influenced by the Hydra-Centaurus
complex. \object{NGC 1365} is a member of Fornax, 
\object{NGC 925} belongs to the
\object{NGC 1023} group and \object{NGC 7331}
 belongs to a small group near the Local
Void. Under these conditions it is not surprising that these
galaxies show deviations from the Hubble law, except that they all
show a tendency to have larger velocity than predicted by the
Hubble law. 

As can be seen from Fig.~\ref{F2}, these galaxies suggest
a much shorter distance to Virgo (for them 
$H_0=75\mathrm{\ km\,^{-1}\, Mpc^{-1}}$ corresponding to
$R_\mathrm{Virgo}=16\mathrm{\ Mpc}$ would be more suitable choice for
the Hubble constant). This means that without further analysis the
data outside the $30\degr$ cone do not allow us to exclude a shorter
distance to Virgo. We examine the effect of a short distance to Virgo
on the galaxies within the $30\degr$ cone in an appendix to this
paper.

Finally, we have used so far systemic velocities, i.e. velocities
not corrected for the virgocentric motions. We tested how
a simple correction would affect Fig.~\ref{F2}. The corrected
velocities shown as crosses in Fig.~\ref{F2} were calculated as
%
\begin{equation}
\label{E16}
\\ V_\mathrm{corr}=V_\mathrm{obs}+[v(d)_\mathrm{H}-v(d)]
\sqrt{1-\sin^2\Theta/d^2}+V_\mathrm{LG}^\mathrm{in}\cos\Theta,
\end{equation}
where $v(d)$ is solved from Eq.~\ref{E10} using parameters for our 
Model 1 and the Hubble velocity
at the distance $d$ measured from the centre of LSC is 
$v(d)_\mathrm{H}=V_\mathrm{cosm}(1)\times d$.
This correction does not resolve the problem.
%
\subsection{The nearest-by galaxies ($R_\mathrm{gal}<4\mathrm{\ Mpc}$)}
\begin{figure}
\resizebox{\hsize}{!}{\includegraphics{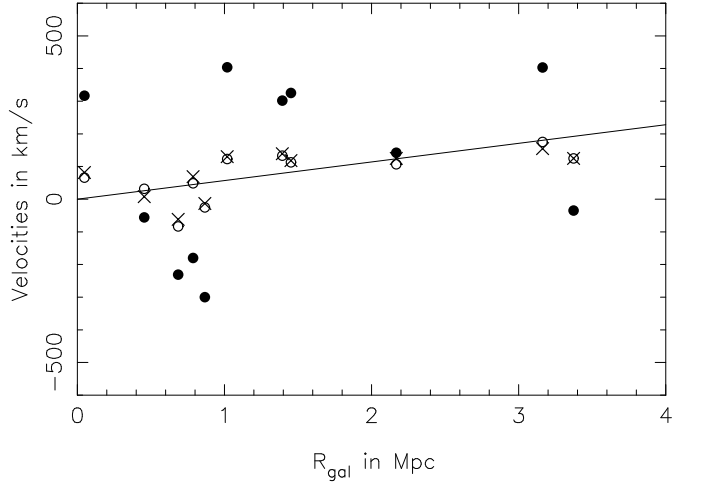}}
\caption{The mean heliocentric velocities (filled circles) and
velocities corrected to the centroid of the LG (according to
Yahil et al. \cite{Yahil77}: crosses; according to Richter
et al. \cite{Richter87}: circles) for the nearest-by galaxies.
}
\label{F3}
\end{figure}

We noticed in the previous subsection that nearest-by galaxies
appear to follow the Hubble law quite well.
In this subsection we take a closer look at these galaxies.
To begin with we note that the 
mean heliocentric velocities $V_\odot$ 
(shown as filled circles in Fig.~\ref{F3}) 
have a rather large
scatter but {\it on the mean} they agree with our primary
choice of $H_0$. It interesting that these very local galaxies
agree with our global value of $H_0$. Furthermore, when
$V_\odot$'s 
are corrected to the centroid of LG one
observes a striking effect. 
In 
Fig.~\ref{F3} one can see how the galaxies after the correction
obey the Hubble law {\it all the way down to}
$R_\mathrm{gal}=0$.
The correction according to
Yahil et al. (\cite{Yahil77}) are shown as crosses and according
to Richter et al. (\cite{Richter87}) as circles. It seems that it
is a matter of taste which correction one prefers. 
%

\section{The virial mass of Virgo}
\subsection{Prediction from Model 1}
Tully and Shaya
(\cite{Tully84}) estimated the virial mass of the Virgo cluster. 
Following 
the notation of Paper I the
mass within $d=0.105$, which corresponds to $\Theta=6\degr$ at
the distance of Virgo, is:
%
\begin{eqnarray}
\label{E17}
 M_\mathrm{virial} &=& 
7.5\times10^{14}M_\odot R_\mathrm{Virgo}/16.8\mathrm{\ Mpc} \nonumber \\
&=& 9.38\times10^{14}M_\odot,
\end{eqnarray}
where the latter equality is based on the adopted distance
$R_\mathrm{Virgo}=21\mathrm{\ Mpc}$. With the 
cosmological parameters ($H_0=57\mathrm{\ km\, s^{-1}\, Mpc^{-1}}$
and $q_0=0.5$) and with our Model 1 
we find using (Eq.~14 of Paper I)
%
\begin{equation}
\label{E18}
\\ M(d) = 14.76\times q_0 h_0^2
\left[\frac{R_\mathrm{Virgo}}{16.8\mathrm{\ Mpc}}\right]^2 \times d^3
\left(1+k'd^{-\alpha}\right),
\end{equation}
where $h_0=H_0/100\mathrm{\ km\, s^{-1}\, Mpc^{-1}}$,
the mass within $d=0.105$ in terms of $M_\mathrm{virial}$:
%
\begin{equation}
\label{E19}
\\ M_\mathrm{pred} = 1.62\times M_\mathrm{virial}.
\end{equation}
The mass deduced agrees with the estimates 1.5-2.0
found by Paper I, where it was suspected that the virial mass estimation
may come from a more concentrated area, which could explain the
higher value obtained from the TB-solution.
\subsection{An alternative solution (Model 3)}
\begin{figure}
\resizebox{\hsize}{!}{\includegraphics{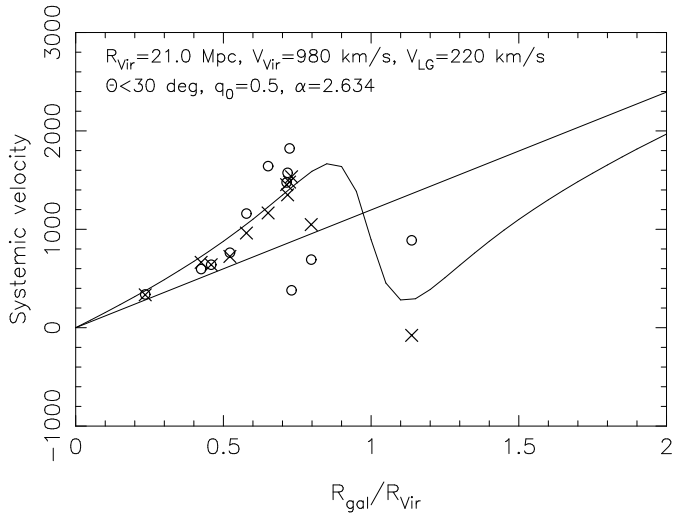}}
\caption{
The systemic velocity vs. distance for galaxies listed
in Table 1 with $\alpha=2.634$ (Model 3) predicting exactly one
Virgo virial mass within a radius $d=0.105$ from the
Virgo. 
Circles refer to the observed velocities and crosses to the
TB-predictions. The straight line is the Hubble law for
$H_0=57\mathrm{\ km\, s^{-1}\, Mpc^{-1}}$ and the
curve is the theoretical TB-pattern for $\Theta=7\degr$.
}
\label{F4}
\end{figure}

We know all the parameters except $\alpha$ needed in Eq.~\ref{E18}.
This led us make a
test by looking for such an $\alpha$ which would bring about exactly one
Virgo virial mass within the radius $d=0.105$ from the Virgo center. 
We found $\alpha=2.634$.
The systemic velocity vs. distance diagram for this model 3 is given in
Fig.~\ref{F4}.
%
Again, \object{NGC 4548} and \object{NGC 4639} 
show anomalous behaviour. Comments
made earlier are valid. There are three other
galaxies showing a relatively large disagreement with the 
given TB-solution. \object{NGC 4535} is close to Virgo at
a small angular distance. Hence the arguments used for \object{NGC 4639}
are acceptable also here. Note that in Paper I this galaxy was
classified as belonging to the region A, where galaxies are
falling directly into Virgo. Furthermore, Federspiel et al.
(\cite{Federspiel98}) classifies this galaxy as a member of the
subgroup B (these galaxies lie within $2.4\degr$ of \object{M49}). This
clearly suggests a distortion in the velocity. \object{NGC 4414} may
belong to the Region B in Paper I. This could explain its low
velocity. It is also rather close to the tangential point so
that projection effects can be important. For \object{NGC 4536} it is
difficult to find any reasonable explanation.
With these notes taken into account we conclude that model 3 
cannot be excluded. 
%
%
%
%
\section{Conclusions}
Studies of large-scale density enhancements in the Local
Universe using spiral field galaxies -- objects suitable for 
such a project -- are discouraged by two factors. The observed distribution
of spirals is rather irregular and the photometric
distances based on Tully-Fisher relation are uncertain due to
large scatter. In the present work we avoided the latter
problem up to a degree by using photometric distances
from the extragalactic $PL$-relation. Such distances are
far more accurate than Tully-Fisher
distances.

It was quite satisfying to find out that the spherically
symmetric TB-model predicts the observed velocity field
well. As a matter of fact, in a recent study Hanski et al.
(\cite{Hanski99}) implemented the TB-model to the mass
determination of the Perseus-Pisces supercluster. Though 
the irregular behaviour of the luminous matter distribution
is even more pronounced in
Perseus-Pisces than in LSC the mass estimates were reasonable.
These findings tend to indicate that the gravitating mass is
more symmetrically distributed than the luminous matter indicates. 

We found a solution (Model 1) predicting a Virgo cluster mass within
$d_{21}=0.105$ of
$M_\mathrm{Virgo}=1.62\times M_\mathrm{virial}$ with
$M_\mathrm{virial}$ given by Tully \& Shaya (\cite{Tully84}).
This result is in agreement with Paper I, where the
difference was suspected to mean that the virial mass
is estimated from a more concentrated volume. Another
plausible explanation is that Virgo is flattened. When presuming
spherical symmetry under such condition  more
mass is required to induce the same effect on the velocities.

We were also able to find a solution predicting exactly one
virial mass. Though this model does not agree with observations
as well as Model 1 the fit is -- considering the large 
uncertainties involved -- acceptable.

We find these results significant in three ways. The intrinsic
behaviour of the sample outside Virgo does not disagree with the 
global value of the Hubble
constant $H_0=57\mathrm{\ km\, s^{-1}\, Mpc^{-1}}$.
The TB-solutions agree with $R_\mathrm{Virgo}=21\mathrm{\ Mpc}$,
a value in excellent concordance with 
$R_\mathrm{Virgo}=20.7\mathrm{\ Mpc}$ given by Federspiel et al.
(\cite{Federspiel98}). 
It is interesting also to be able to predict exactly one Virgo
virial mass in the Einstein-deSitter universe, because in
Paper I the mass predictions were larger than one and because
$\Omega_0=1$ is strongly supported by theoretical
considerations (both Inflation and Grand Unification require this value).
Discussion on the cosmological constant $\Lambda$ is postponed
to a later phase of our research programme.  

Having said all this one should be cautious. Results presented
in this paper are based on frontside galaxies only. As can be
seen from Fig.~4. in Paper I, predicted systemic velocities are
more sensitive to the model parameters in the backside than in the
front. In the next paper of this series we intend to find a few
background galaxies with as reliable distance moduli as possible from other
photometric distance estimators, in particular the Tully-Fisher
relation with proper care taken of the selection effects.
%
%
%
%
\begin{acknowledgements}{
This work has been partly supported by the Academy of Finland
(project 45087: ``Galaxy Streams and Structures in the nearby
Universe" and project ``Cosmology in the Local Universe").
We have made use of the Lyon-Meudon Extragalactic Database LEDA
and the Extragalactic Cepheid Database. Finally we are 
grateful for the referee for constructive criticism and suggestions
which have helped us to improve this paper.} 
\end{acknowledgements}  
%
%
\appendix
\section{Is $R_\mathrm{Virgo}\approx16\mathrm{\ Mpc}$?}
In the present paper we fixed the distance to the centre of
LSC by adopting a $H_0$ from external considerations and
by fixing the cosmological velocity of the centre of LSC
leading to $R_\mathrm{Virgo}=21\mathrm{\ Mpc}$.
In Sect. 5.1 we noted that the galaxies outside the $30\degr$
cone measured from the centre of LSC do not allow us to
exclude higher values of $H_0$. Due to the fixed velocity
these higher values, if really cosmological, 
necessarily lead to a shorter distance
to the centre of LSC.
%
\begin{figure}
\resizebox{\hsize}{!}{\includegraphics{fig5.epsi}}
\caption{As Fig.~\ref{F1} but now the distance to Virgo
is $R_\mathrm{Virgo}=16\mathrm{\ Mpc}$.
}
\label{FA1}
\end{figure}

In this appendix we examine what happens to the TB pattern
within the $30\degr$ cone where the TB model is relevant if
we use $R_\mathrm{Virgo}=16\mathrm{\ Mpc}$. The result is
shown in Fig.~\ref{FA1}. The infall pattern is still clearly
present. The agreement between the observed points (circles)
and the predicted points (crosses) is not as good as in
Figs.~\ref{F1} and~\ref{F4}. 
When Fig.~\ref{FA1}
is carefully compared with Fig.~\ref{F1} one notes that
in Fig.~\ref{FA1} the observed systemic velocities are 
{\it systematically} smaller than the predicted ones. This
is because for the shorter Virgo distance galaxies in front get closer
to Virgo and thus the dynamical influence of Virgo should be larger.
In Fig.~\ref{F1} we observe no such systematic effect. 
It is thus
possible to say that our sample is more favourable to a long
distance scale than to a short scale.
 
\begin{figure}
\resizebox{\hsize}{!}{\includegraphics{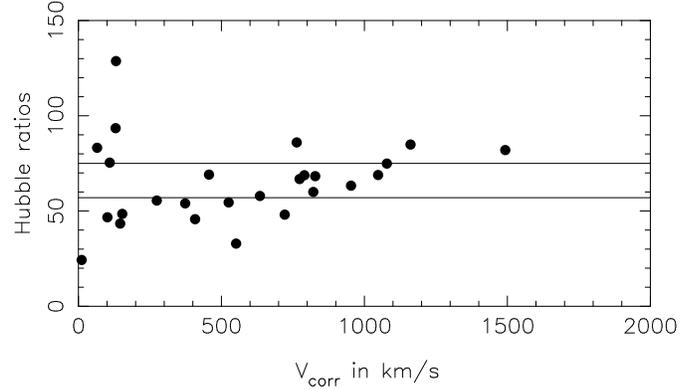}}
\caption{The Hubble ratios as a function
of the corrected velocity for $R_\mathrm{Virgo}=21\mathrm{\ Mpc}$. 
}
\label{FA2}
\end{figure}

We also examine the behaviour of the Hubble ratios 
$V_\mathrm{corr}/R_\mathrm{gal}$
as a function of $V_\mathrm{corr}$ (cf. Eq.~\ref{E16}).
In Fig.~\ref{FA2} the correction is based on 
$R_\mathrm{Virgo}=21\mathrm{\ Mpc}$ and in Fig.~\ref{FA3} on 
$R_\mathrm{Virgo}=16\mathrm{\ Mpc}$. 
The two values of $H_0$ used in Fig~\ref{F2} 
are also given as straight horizontal lines.
In both diagrams we observe
increase in the Hubble ratios as $V_\mathrm{corr}$ 
increases starting at $V_\mathrm{corr}\approx600\mathrm{\ km\, s^{-1}}$.
In particular, for the shorter Virgo distance this effect is
quite pronounced. 
Sandage and Tammann
have on many occasion stressed that the value of the Hubble constant,
$H_0$, should not increase without any clear physical reason. In
the absence of such the increase in $H_0$ is a warning signal that 
there is a bias present (e.g. the Malmquist bias). 
This is also our tentative interpretation
for the increase in $H_0$.
One should also pay attention to the fact that in both cases
the mean Hubble ratio is $\sim55\mathrm{\ km\, s^{-1}\, Mpc^{-1}}$
for galaxies below $V_\mathrm{corr}=600\mathrm{\ km\, s^{-1}}$.

However, part of the apparent increasing trend in $H_0$ 
could be an artefact of the velocity dispersion. If we
overestimate the velocities galaxies have a tendency to move upwards
and right. 
What is the nature of the suspected bias and its relation to the
incompleteness bias of Lanoix et al. (\cite{Lanoix99c}) as well
as what is the influence of the velocity effect
are amongst the topics we discuss in the next paper in this series.    
%
%
%
%
\begin{figure}
\resizebox{\hsize}{!}{\includegraphics{fig7.epsi}}
\caption{As Fig.~\ref{FA2} but now 
$R_\mathrm{Virgo}=16\mathrm{\ Mpc}$.
}
\label{FA3}
\end{figure}
%
%

%
%
\end{document}